\documentclass[10pt,conference]{IEEEtran}
\IEEEoverridecommandlockouts
\usepackage{cite}
\usepackage{amsmath,amssymb,amsfonts}
\usepackage{algorithmic}
\usepackage{graphicx}
\usepackage{textcomp}
\usepackage{xcolor}

\usepackage{booktabs}
\usepackage{multirow}
\usepackage{makecell}
\usepackage{subfigure}
\usepackage{tcolorbox}
\usepackage{enumitem}
\usepackage{float}
\usepackage{tabularx}
\usepackage{url}
\usepackage[breaklinks]{hyperref}

\newcommand{\RNum}[1]{\uppercase\expandafter{\romannumeral #1\relax}}

\definecolor{lgray}{gray}{0.9}

\def\BibTeX{{\rm B\kern-.05em{\sc i\kern-.025em b}\kern-.08em
    T\kern-.1667em\lower.7ex\hbox{E}\kern-.125emX}}
\begin{document}

\title{Instruct or Interact? Exploring and Eliciting LLMs’ Capability in Code Snippet Adaptation Through Prompt Engineering}

%
% \author{\IEEEauthorblockN{Anonymous Author(s)}}
\author{
\IEEEauthorblockN{Tanghaoran~Zhang, Yue~Yu\textsuperscript{*}\thanks{\textsuperscript{*}Yue Yu is the corresponding author.}, Xinjun~Mao, Shangwen~Wang, Kang~Yang, Yao~Lu, Zhang~Zhang and Yuxin~Zhao}
\thanks{Tanghaoran~Zhang, Xinjun~Mao, Kang~Yang, Yao~Lu, Zhang~Zhang and Yuxin~Zhao are with the Key Laboratory of Software Engineering for Complex System.}
\IEEEauthorblockA{College of Computer Science and Technology \\
National University of Defense and Technology \\
Changsha, China \\
\{zhangthr, yuyue, xjmao, wangshangwen13, yangkang, luyao08, zhangzhang14, yuxinzhao\}@nudt.edu.cn}
}

\maketitle

\begin{abstract}
Code snippet adaptation is a fundamental activity in the software development process. Unlike code generation, code snippet adaptation is not a ``free creation'', which requires developers to tailor a given code snippet in order to fit specific requirements and the code context. Recently, large language models (LLMs) have confirmed their effectiveness in the code generation task with promising results. However, their performance on code snippet adaptation, a reuse-oriented and context-dependent code change prediction task, is still unclear.
To bridge this gap, we conduct an empirical study to investigate the performance and issues of LLMs on the adaptation task. We first evaluate the adaptation performances of three popular LLMs and compare them to the code generation task. Our result indicates that their adaptation ability is weaker than generation, with a nearly 15\% decrease on pass@1 and more context-related errors. By manually inspecting 200 cases, we further investigate the causes of LLMs’ sub-optimal performance, which can be classified into three categories, \emph{i.e.,} \textit{Unclear Requirement}, \textit{Requirement Misalignment} and \textit{Context Misapplication}.
Based on the above empirical research, we propose an interactive prompting approach to eliciting LLMs' ability on the adaptation task. 
Specifically, we enhance the prompt by enriching the context and decomposing the task, which alleviates context misapplication and improves requirement understanding. Besides, we enable LLMs' reflection by requiring them to interact with a human or a LLM counselor, compensating for unclear requirement.
Our experimental result reveals that our approach greatly improve LLMs' adaptation performance. The best-performing Human-LLM interaction successfully solves 159 out of the 202 identified defects and improves the pass@1 and pass@5 by over 40\% compared to the initial instruction-based prompt. Considering human efforts, we suggest multi-agent interaction as a trade-off, which can achieve comparable performance with excellent generalization ability. We deem that our approach could provide methodological assistance for autonomous code snippet reuse and adaptation with LLMs.
\end{abstract}


\section{Introduction}
\label{sec1:intro}
With the thriving growth of the open-source community, software reuse is widely adopted to efficiently deliver high-quality software products. Besides component-based reuse, reusing online code snippets has become a common practice in modern software development~\cite{Yang2017,Baltes2019,Manes2021,Huang2022,wang2023natural,wang2023two}. These public available snippets from various platforms, e.g., GitHub and Stack Overflow, are widely reviewed and proofed by a large number of open-source contributors. Compared to writing the code from scratch, leveraging the recognized knowledge to build software has lower cost and less risks~\cite{Brandt2009,Yang2016,wang2024fusing}.
However, online code snippets often fail to meet the specific needs of developers. Hence, apart from simple copy-and-paste, developers are usually required to adapt these code snippets according to their development contexts to ensure the correctness and maintainability of the code~\cite{Zhang2019,Mondal2019,Zhang2024}. Over the years, several code snippet adaptation techniques and tools have been proposed to facilitate this daily activity~\cite{Cottrell2008,Wightman2012,Reid2020,Terragni2021}, but it is still a pending issue for its automation.

Recent advancements in artificial intelligence have been marked by the emergence of large language models (LLMs), such as ChatGPT~\cite{OpenAI2023}. These LLMs are distinguished by their large scale of parameters and emergent abilities~\cite{Brown2020,Wei2022} in processing natural language, which stems from extensive training on diverse data sources. This training empowers them with abilities applicable to numerous software engineering tasks, e.g., code generation~\cite{Gao2023,Du2023}, code summarization~\cite{geng2024large}, and automated program repair~\cite{Sobania2023,Cao2023,Gao2023,Xia2023,qin2024agentfl}. Code snippet adaptation can also be performed with LLMs' generation ability when the task is properly described. Therefore, it is imperative to investigate the potential of LLMs to perform adaptation.

LLMs are utilized through the ``Pre-train, Prompt, and Predict'' paradigm~\cite{Liu2021}, which requires users to engage with them through a set of textual inputs, i.e., prompt. Prompts enable us to teach LLMs unseen tasks with no need for fine-tuning them or modifying their architectures, known as programming in natural language~\cite{Reynolds2021}. For instance, Alice could ask LLMs to debug her code by simply entering ``\textit{Please help me find the bugs in the following code: $<$code$>$}'' as an input prompt. Besides, the prompt selection could significantly influence the capabilities of LLMs~\cite{Liu2021,Reynolds2021,Rodriguez2023}. Therefore, crafting appropriate prompts is the way of eliciting the ability of LLMs to perform the corresponding task, which is known as ``prompt engineering''~\cite{Reynolds2021,Liu2022}. However, it is challenging to design an optimal prompt for code snippet adaptation. The reason is that performing adaptations requires an accurate understanding of its context. LLMs should be instructed to identify and strictly adhere to the constraints of the context, e.g., dependencies on specific fields and methods, like ``dancing in the fetters''.

Therefore, we focus on the prompt engineering of code snippet adaptation to explore and elicit LLMs' capabilities. 
To this end, we first conduct an empirical study to investigate the effectiveness of LLMs and their limitations on code snippet adaptation. Specifically, we evaluate the adaptation performances of three popular LLMs based on the \textit{ClassEval} benchmark~\cite{Du2023}. The results show that the best-performing LLM, i.e., GPT-3.5, achieves 52.34, 60.98, and 47.05 on pass@1, pass@5, and CodeBLEU. However, their adaptation performance is inferior (15\% in pass@1) to generation. Therefore, we further investigate the issues of LLMs' adaptations on 200 sampled cases adapted by GPT-3.5. For each case, we inspect the failed test cases and identify defects from the adapted code. We find that adaptation includes more context-related errors than generation, indicating LLMs' unawareness of the context. Then we annotate the origins and causes of above defects. 74\% of them are pre-existent and overlooked by LLMs and significantly fewer adaptations are made actually than required, highlighting LLMs' laziness on adaptation. Besides, we summarize three categories of the root causes that lead to LLMs' failures, including \textit{Unclear Requirement}, \textit{Requirement Misalignment} and \textit{Context Misapplication}.

Motivated by our observations, we propose an interactive prompting approach to addressing identified LLMs' issues in adaptation. Our approach enriches the prompt with more information to avoid LLMs' context misapplication. It decomposes the adaptation task with a multi-turn conversation style to alleviate LLMs' burden of understanding. Then we integrates an interaction workflow, allowing LLMs to flip their roles to refine the requirements by asking questions. The interaction is implemented with two schemes, one through a human-in-the-loop supervision and the other through a multi-agent collaboration. The interaction compensates for unclear requirements and enables LLMs' reflection to identify their confusion. The result demonstrates that Human-LLM interaction achieves better performances, which solves 159 out of the 202 previously-identified defects and elevates the performance by 41.4\%, 42.6\%, and 26.1\% on pass@1, pass@5, and CodeBLEU. However, human interventions may incur labor costs. As a trade-off, our proposed Multi-Agent interaction could achieve comparable performance (only about 5\% decrease in pass@5) and also have promising generalization ability on both conversational and instruction-tuned LLMs.

In summary, this paper makes the following contributions:

\begin{itemize}[leftmargin=*]
    \item To the best of our knowledge, we conduct the first study to evaluate the effectiveness of LLMs on the code snippet adaptation task. Our results demonstrate that the adaptation task is more challenging for LLMs than generation.
    \item We find that LLMs are lazy adaptation makers and have weak context awareness. Furthermore, we summarize three categories of nine causes of their failures in adaptation, revealing their current limitations. 
    \item We propose a novel interactive prompting approach to utilizing LLMs' in code snippet adaptation. It provides an effective use of LLMs in software reuse tasks, supporting current reuse-oriented software engineering methodology.
    \item  Our accessible source code and annotated data\footnote{https://github.com/ztwater/Instruct-or-Interact.} will facilitate the replication and application of our study.
\end{itemize}

The rest of this paper is organized as follows:
Section~\ref{sec2:empirical study} describes our empirical study.
Section~\ref{sec3:approach} introduces our interactive prompting approach.
Section~\ref{sec4:evaluation} evaluates our proposed approach. 
% Section~\ref{sec5:discussion} discusses the current limitations of prompting LLMs and our suggestions. 
Section~\ref{sec5:rw} discusses the related work of this study.
Section~\ref{sec6:threats} presents threats to validity, and Section~\ref{sec7:conclusion} concludes the paper.

\begin{figure*}[h]
    \centerline{\includegraphics[width=\linewidth]{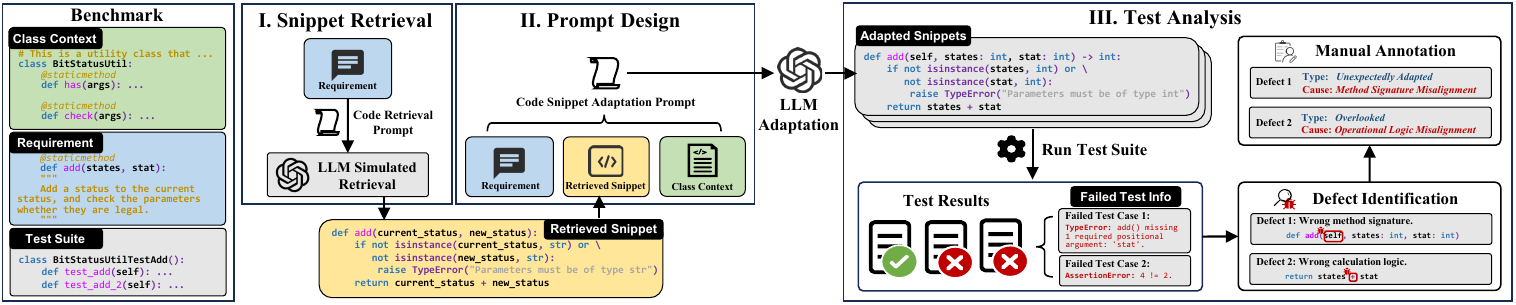}}
    \caption{The framework of our empirical study.}
    \label{fig:framework}
\end{figure*}

\section{Empirical Study}
\label{sec2:empirical study}
\subsection{Research Questions}
To evaluate the performance and issues of LLMs in code snippet adaptation, we structure the goal of our empirical study in the following research questions.

\textbf{\textit{RQ1: How effective are LLMs on the adaptation task?}}
Existing studies have evaluated LLMs' performance on a wide spectrum of software engineering tasks. However, their effectiveness on the code snippet adaptation task remains unexplored. To this end, we investigate the adaptation performance of three widely used LLMs' in different settings. 

\textbf{\textit{RQ2: What are the current issues of LLMs' adaptations?}}
To better utilize LLMs for code snippet adaptation, it is necessary to understand their current limitations from their adaptation results. To this end, we analyze the test results of the adapted code and their failure-inducing defects by inspecting 1,000 GPT-3.5 adapted snippets in 200 cases.

\textbf{\textit{RQ3: What are the root causes of LLMs' adaptation failures?}}
This RQ aims to understand why LLMs fail to adapt snippets to their context. To achieve this goal, we conduct a thematic analysis to summarize the underlying reasons of LLMs' failures. The answer is of great importance to understanding their nature of performing adaptations.

\subsection{Study Design}
\label{sec2.2:study}
Our empirical study follows a mixed-method research methodology. Fig.~\ref{fig:framework} describes its framework. As there is no benchmark for code snippet adaptation, we first supplement a code generation benchmark, \textit{ClassEval}, with a simulated snippet retrieval process, in which a group of retrieved snippets are generated as the adaptation source. The core task for LLMs is to adapt these snippets correctly to their class context. To this end, we design a code snippet adaptation prompt to instruct LLMs to complete the task. Finally, through quantitative and qualitative test analysis, we evaluate LLMs' adaptation performance and summarize their issues and root causes from failed cases.

\begin{figure}[h]
    \centering
    \includegraphics[width=\linewidth]{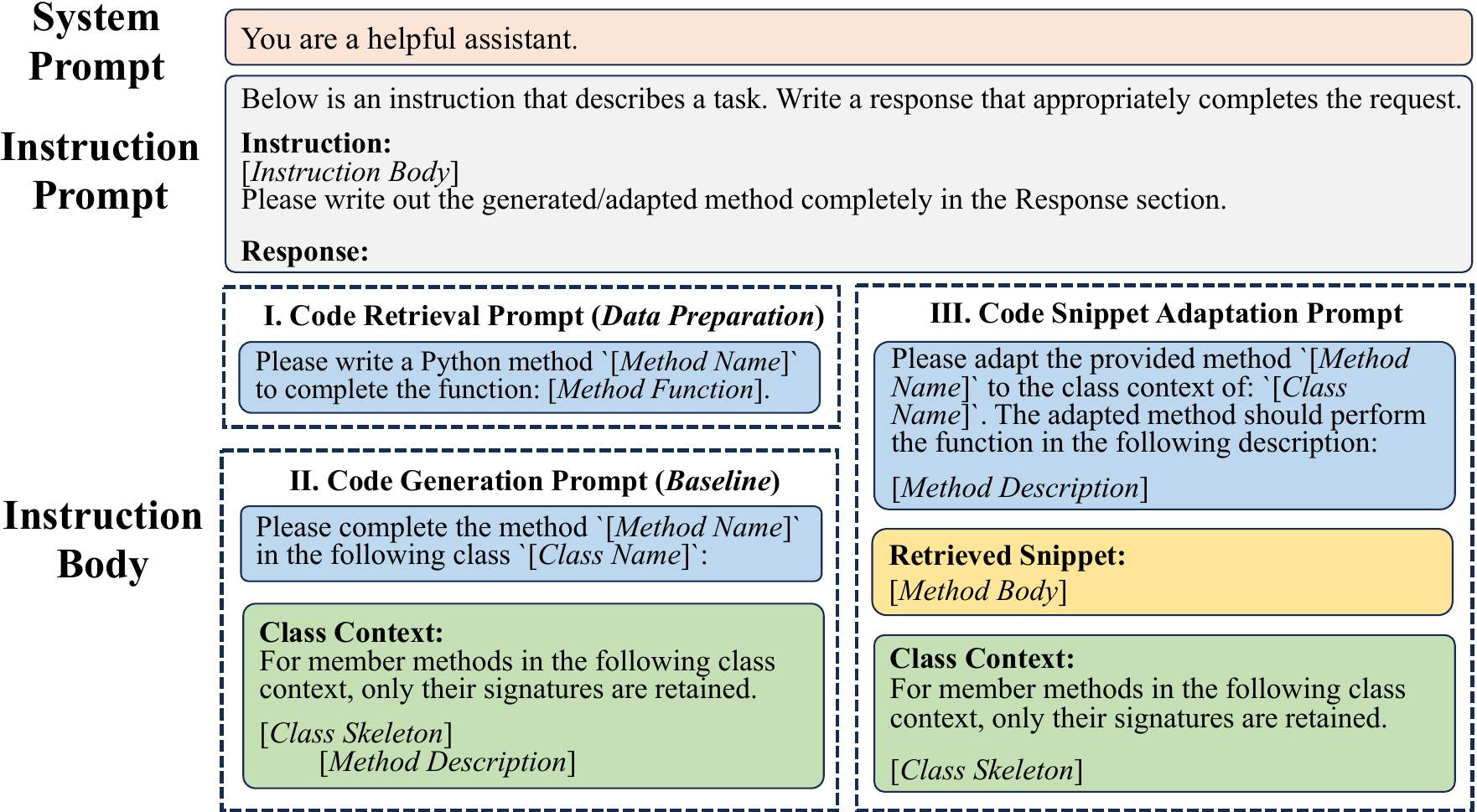}
    \caption{The prompt design of code retrieval, code generation and code snippet adaptation.}
    \label{fig:prompt}
\end{figure}

\textbf{Snippet Retrieval.}
To support the evaluation of code snippet adaptation, we supplement \textit{ClassEval} with retrieved snippets. Typically, developers retrieve the snippets for reuse by searching and selecting from online open-source communities. However, this process is traditionally labor-intensive, requiring precise query formulation and answer selection. To streamline this, we leverage GPT-3.5 to simulate snippet retrieval by only providing the method name and its description as requirement. For each method in the benchmark, we obtain a ``general'' snippet as the source and LLMs will further adapt it to the specific class context. Our prompt used in snippet retrieval is discussed below and shown in Fig.~\ref{fig:prompt}.

\textbf{Prompt Design.} In line with the established practices for instructing LLMs~\cite{Luo2023,Du2023}, we develop the prompt skeleton comprising two parts: 1) a general system prompt and 2) an instruction prompt detailing the specific requirements for the task. Our prompts for code retrieval (data preparation), code generation (baseline), and code snippet adaptation follow this two-part structure and are differentiated by their instruction contents, as shown in Fig.~\ref{fig:prompt}. For code retrieval, its prompt includes only the task description without additional context, while the code generation prompt also incorporates class context and the target method description at the end. The adaptation prompt further includes the retrieved snippet. Our design aligns with current guidelines~\cite{Zhao2023}, which advocate for clear separation of different prompt components using model-friendly markers like ``\#\#\#''. This approach not only aids in clarity but also enhances LLMs' ability to process the prompts.

\textbf{Test Analysis.}
We perform manual test analysis to identify the issues and root causes of LLMs' adaptations based on the \textit{ClassEval} benchmark~\cite{Du2023} (see Section~\ref{sec2.3:setting}). To this end, we randomly select 200 of 410 methods, which exceeds the sample size required for a 95\% confidence level and a 5\% margin of error. Leveraging \textit{ClassEval}'s test suites, the first and fifth authors identify defects in the adapted methods from the failed test cases. Given that defects are not necessarily introduced by adaptations, e.g.,
the retrieved snippet does not satisfy the requirement, we annotate all identified defects into three types based on their origins:
1) \textit{Overlooked}: The defect exists before adaptation and has not been adapted by LLM; 
2) \textit{Invalidly Adapted}: The defect exists before adaptation, and it is adapted incorrectly by LLM;
3) \textit{Unexpectedly Adapted}: The defect emerges after adaptation due to incorrect LLM's adaptations.
Disagreements are initially discussed between the two annotators, and unresolved cases are further reviewed by the second author for a majority decision. Eventually, the identified defects and their types are approved by all three annotators. The Cohen's Kappa value~\cite{Cohen1960} of the classification is 0.88, which indicates a strong agreement between annotators. 

Moreover, we employ thematic analysis~\cite{Braun2006,Berg2017} to investigate root causes of our identified defects. The same two authors independently annotate all sampled methods with their root cause descriptions. This labeling process focuses on discrepancies between the adapted snippets and canonical solutions. Subsequently, the annotators re-read and group the codes with similar descriptions and generate the initial themes for their root causes. Then they iteratively re-evaluate and group the themes until they are established. The final themes are consolidated and refined by the authors, ensuring they accurately reflect the underlying issues.

To support our annotation, we build an interface for each adaptation, which displays the original requirement, reused code snippet, LLM-adapted code, canonical solution, and test results. All participants in this study have over five years of Python programming experience. Eventually, we obtain three categories of causes, i.e., \textbf{Unclear Requirement}, \textbf{Requirement Misalignment}, and \textbf{Context Misapplication}, which are further explained in Section~\ref{sec:causes}.

\subsection{Experiment Setting}
\label{sec2.3:setting}
\textbf{Dataset.} Our study utilizes the \textit{ClassEval} benchmark~\cite{Du2023} to evaluate LLMs' adaptations. \textit{ClassEval}, originally a class-level code generation benchmark comprising 100 Python classes and their associated test suites, is apt for our context-dependent adaptation task. 
Besides, \textit{ClassEval} is a manually crafted dataset covering diverse topics, which simulates real software development scenarios. It was released in 2023 and hence could mitigate the training data exposure for LLMs. Based on \textit{ClassEval}, we obtain the class-level context by excluding the target method and all other methods dependent on it (i.e., caller methods) for each class. Finally, we derive 410 method-level adaptation cases by pairing our retrieved snippets with their corresponding context. 

\textbf{Studied Models.} We select three widely studied LLMs, including one conversational LLM (GPT-3.5~\cite{OpenAI2023}) and two instruction-tuned code LLMs (CodeLlama~\cite{Roziere2023}, Llama-3~\cite{Meta2024}). These LLMs are chosen for their capability to interpret natural language prompts, which allows us to instruct them for code snippet adaptation. For GPT-3.5, we use \textit{gpt-3.5-turbo-0613} checkpoint~\cite{OpenAI2023} through the OpenAI API. As for instruction-tuned LLMs, we select the 7B, 13B and 34B versions for CodeLlama and the 8B version for Llama-3.

\textbf{Baseline.} As there is not a general-purpose predictive approach for code snippet adaptation, we utilize code generation with LLM as our baseline to evaluate the current situation of LLMs' adaptation. The prompt used is shown in Fig.~\ref{fig:prompt}. Our design ensures a fair comparison between code generation and code snippet adaptation prompts by providing the same level of details in the context according to the taxonomy outlined by Santu et al.~\cite{Santu2023}. Their difference is that adaptation requires LLMs to perform the task based on the retrieved snippet.

\textbf{Evaluation Metrics.} We use the pass@$k$ as our primary metric to evaluate the correctness of the generated adaptations by LLMs. It measures the likelihood of problem-solving within $k$ attempts based on unit test performance.
\begin{equation}
    {\rm pass}@k=\mathbb{E}_{{Problems}}\left[1-\frac{\binom{n-c}{k}}{\binom{n}{k}}\right]
\end{equation}
\noindent where $k$ represents the number of candidates sampled from all results with the total number of $n$. $c$ is the number of samples that pass the tests. Considering the generation cost, we set $n$ to 5 in line with previous work \cite{Du2023}. Specifically, we calculate pass@$1$ and pass@$5$ for each adaptation case, reflecting both the accuracy and practical solution-finding efficiency, i.e., whether a solution can be found by skimming through at most 5 snippets generated by LLMs.

Furthermore, we also adopt CodeBLEU~\cite{Ren2020} as a complementary evaluation metric. It considers the n-gram match and syntactic and semantic similarity via abstract syntax trees (ASTs) and data flow. We use it to evaluate similarities between generated snippets and canonical solutions, offering insights beyond mere test pass rates.

\textbf{Implementation Details.}
To alleviate the impact of LLMs' response randomness, we employ nucleus sampling~\cite{Holtzman2020} according to previous studies~\cite{Du2023}. For each task, five responses are randomly generated and collected for evaluation. Our experiments also explore the impact of temperature settings (0 to 1, at 0.2 intervals) on LLMs' performance. We set the maximum token window to 2,048 for all LLMs, ensuring consistent context lengths. All experiments are conducted with two GeForce RTX 4090-24G GPU.

\subsection{RQ1: LLMs' Adaptation Performance}

\begin{table*}[htbp]
    \caption{Adaptation Performance of LLMs in Different Temperature Settings}
    \scriptsize
    \setlength{\tabcolsep}{4.5pt}
    \begin{center}
        \begin{tabular}{|c|ccc|ccc|ccc|ccc|ccc|ccc|}
            \Xhline{2\arrayrulewidth}
            \multirow{2}{*}{\textbf{Model}} & \multicolumn{3}{c|}{Temperature=0} & \multicolumn{3}{c|}{Temperature=0.2} & \multicolumn{3}{c|}{Temperature=0.4} & \multicolumn{3}{c|}{Temperature=0.6} & \multicolumn{3}{c|}{Temperature=0.8} & \multicolumn{3}{c|}{Temperature=1} \\
            \cline{2-19}
            & p@1 & p@5 & CB & p@1 & p@5 & CB & p@1 & p@5 & CB & p@1 & p@5 & CB & p@1 & p@5 & CB & p@1 & p@5 & CB \\
            \hline
            GPT-3.5 &
            51.17 & 51.95 & \textbf{47.51} &
            51.37 & 55.85 & 47.50 &
            51.90 & 57.80 & 47.45 &
            51.90 & 57.56 & 47.23 &
            \textbf{52.34} & \textbf{60.98} & 47.05 &
            51.46 & 60.24 & 46.80 \\
            \hline
            CodeLlama-7B &
            39.61 & 39.76 & 42.13 &
            37.61 & 45.37 & 41.87 &
            37.22 & 48.54 & 41.57 &
            34.10 & 47.32 & 40.39 &
            32.10 & \textbf{49.51} & 38.90 &
            29.95 & 49.02 & 36.95 \\
            CodeLlama-13B &
            34.24 & 34.63 & 40.30 &
            34.20 & 44.63 & 40.12 &
            33.51 & 48.54 & 39.24 &
            30.49 & 48.78 & 36.98 &
            26.63 & 47.07 & 34.52 &
            22.98 & 47.07 & 31.36  \\
            CodeLlama-34B &
            \textbf{40.10} & 40.49 & \textbf{43.12} &
            39.17 & 46.83 & 42.89 &
            36.00 & 47.80 & 40.27 &
            32.78 & \textbf{49.51} & 37.56 &
            28.63 & 49.02 & 34.24 &
            24.83 & 48.29 & 31.88 \\
            \hline
            Llama-3-8B &
            \textbf{45.61} & 46.59 & 45.80 &
            45.32 & 52.68 & \textbf{45.92} &
            43.80 & 55.37 & 45.72 &
            42.24 & \textbf{58.29} & 45.34 &
            41.27 & 57.80 & 45.06 &
            39.07 & 55.61 & 44.52 \\
            \hline
            GPT-3.5-Gen &
            66.49 & 67.07 & \textbf{55.91} &
            66.98 & 70.98 & 55.87 &
            \textbf{67.07} & 73.17 & 55.56 &
            66.93 & 74.15 & 54.47 &
            65.27 & \textbf{75.61} & 54.23 &
            64.49 & 73.66 & 53.46 \\
            \Xhline{2\arrayrulewidth}
        \end{tabular}
    \end{center}
    \label{tab:perf-base}
\end{table*}

% 1.1 basic evaluation
Table \ref{tab:perf-base} illustrates the performance of our selected LLMs in different temperature settings and parameter sizes. For the adaptation task, GPT-3.5 performs the best, achieving 52.34 and 60.98 in pass@1 and pass@5 respectively. Llama-3-8B ranks the second with a competitive pass@5 score 58.29 when the temperature is 0.6, but its pass@1 is much lower than GPT-3.5. The performance of CodeLlama on all parameter sizes are similar and inferior to the above two LLMs. Its best pass@1 and pass@5 are 40.10 and 50.24. Specifically, CodeLlama-34B achieves the best pass@1 and CodeBLEU scores, while its 7B version behaves better when the temperature rises. 

As for the temperature parameter, its optimal setting of GPT-3.5 is 0.8 on the adaptation task, as evidenced by a marginal improvement in pass@1 and a significant improvement in pass@5 ($p<0.001$, Mann-Whitney U test~\cite{Mann1946}). For instruction-tuned LLMs, there is a similar trend. The lower temperature leads to better pass@1 and CodeBLEU scores. A slightly higher temperature, e.g., 0.6 or 0.8, improves the pass@5 score. The reason behind is that a higher temperature introduces more variability in LLMs' outputs, facilitating a broader exploration of potential adaptations. 
In contrast, a temperature setting of 0 stabilizes the generated code structure, resulting in higher CodeBLEU scores. In all, we adopt the optimal temperature setting, i.e., 0.8, for GPT-3.5 in the following experiments.

Taking the best performing LLM (i.e., GPT-3.5) as an example, its performance in code snippet adaptation exhibits a decrease of 14.73, 14.63, and 8.40 in pass@1, pass@5, and CodeBLEU compared to that of code generation. As introduced in Section \ref{sec2.2:study}, their prompts contain the same contextual information. The reason behind the performance gap may be that GPT-3.5 is a generative model designed for predicting the next token, making it well-suited for code generation from scratch. However, adaptation tasks require the comprehension of multiple fragments of a compound corpus and context-based modifications to existing code snippets. This poses a challenging task for current generative LLMs.

In addition to the model type, size and its temperature setting, which can be seen as factors from the model aspect, factors from the task aspect may also impact LLMs' adaptation performance, e.g., the complexity of the task. In our study, we use the \textit{adaptation size}, a.k.a., the number of the required AST edits from retrieved snippet to the canonical solution as the measure for task complexity.

\begin{figure}[h]
    \centerline{\includegraphics[width=\linewidth]{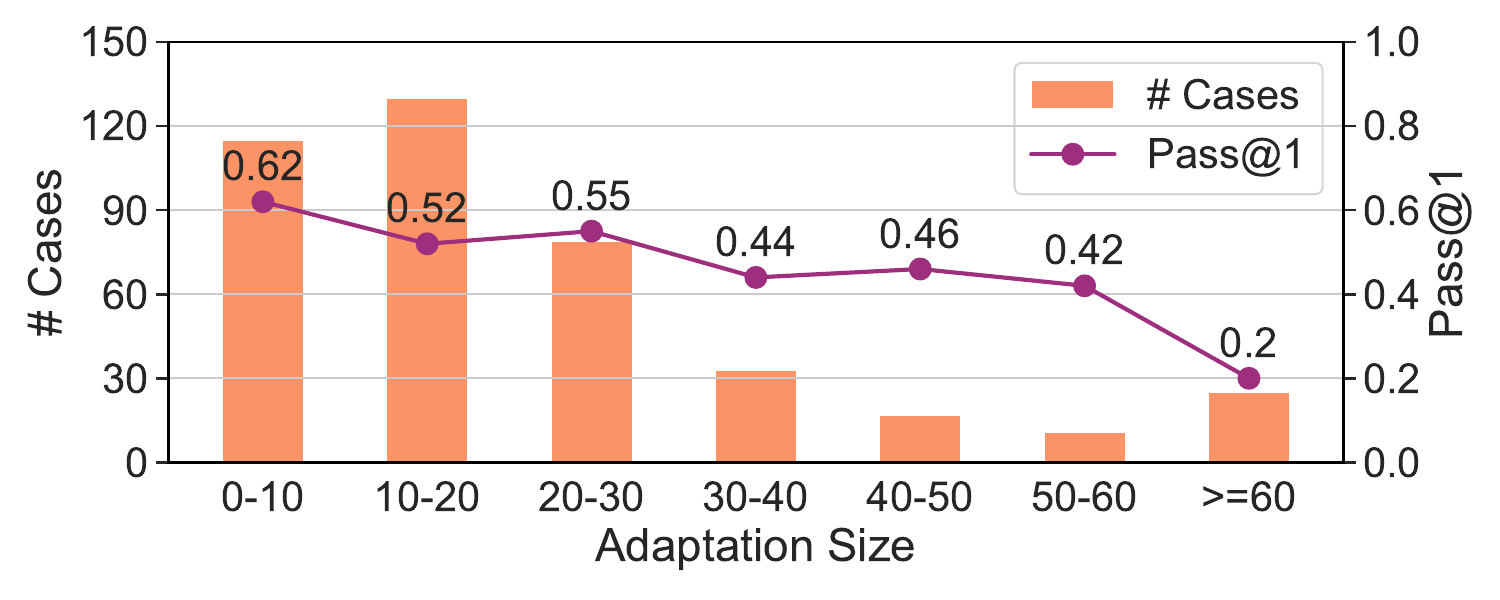}}
    \caption{The distribution of the adaptation size and its relationship with the pass@1 score of GPT-3.5 model on the adaptation task.}
    \label{fig:pass1-size}
\end{figure}

Fig.~\ref{fig:pass1-size} illustrates the distribution of the adaptation size and its relationship with the pass@1 score of GPT-3.5. All adaptation cases are grouped into seven by the adaptation size with an interval of 10. The majority of the cases have the adaptation size below 30. There is a decreasing trend of GPT-3.5's pass@1 with the rising of adaptation size. Two obvious drops in the pass@1 occur when the adaptation size exceeds 30 and 60. The trend indicates that LLMs' ability to accurately complete adaptations is negatively correlated with the scale of adaptations.

\begin{center}
    \begin{tcolorbox}[colback=lgray,colframe=black,width=\linewidth,arc=1mm,boxrule=1pt,top=2pt,bottom=2pt,left=3pt,right=3pt]
        \textbf{Finding 1:} GPT-3.5 with the temperature of 0.8 achieves the best adaptation performance. However, there is still a 15\% gap compared to generation. Besides, LLMs' adaptation performance decreases when the adaptation size rises.
    \end{tcolorbox}
\end{center}

\subsection{RQ2: Issues in LLMs' Adaptations}
To obtain an in-depth understanding of the issues of LLMs' sub-optimal adaptation performance, we perform test analysis on the results of the most advanced model, GPT-3.5, with its best settings in RQ1.
% Based on the best model and parameter setting from RQ1, i.e., GPT-3.5 and $temperature=0.8$, we randomly select 200 methods (each as an adaptation case with five adapted method instances) and further explore the problems leading to LLMs' sub-optimal adaptation performance from the test results. 
Among 200 selected cases, there are a total of 87 cases where all five adapted snippets passed all the tests. There are 35 cases where at least one adaptation passed the tests and 78 cases where all adaptations failed.

\begin{figure}[h]
    \centerline{\includegraphics[width=\linewidth]{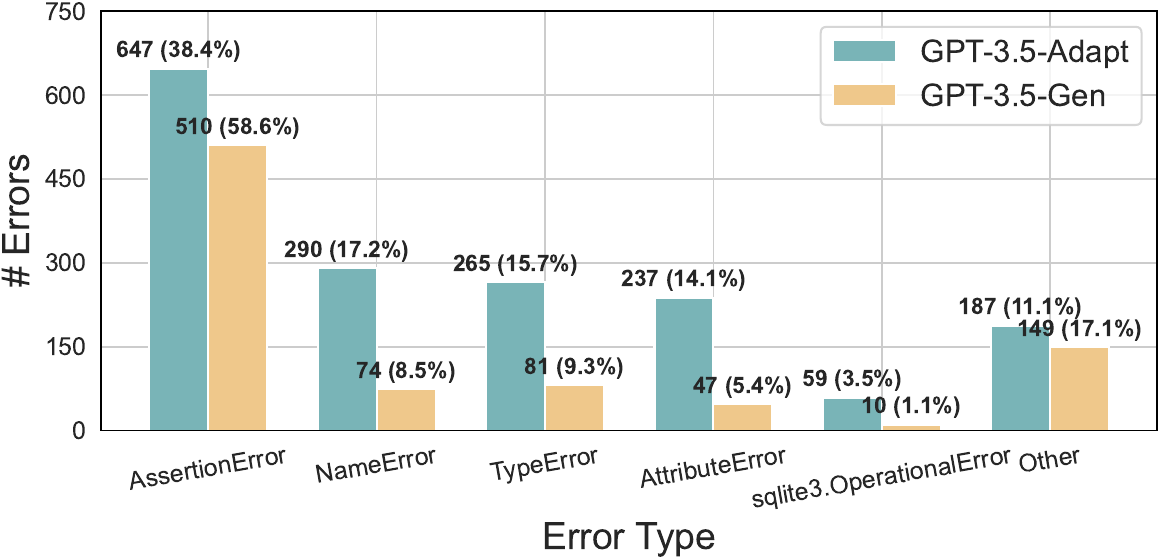}}
    \caption{The distribution of test error types of GPT-3.5's adaptation (1,685 in total) and generation (871 in total). Errors with fewer occurrences are in the Other category.}
    \label{fig:test-err}
\end{figure}

We first compare the distribution of test errors during the adaptation to that in generation, as depicted in Fig.~\ref{fig:test-err}. The total number of errors in the adaptation and generation task are 1,685 and 871, respectively. The most prevalent error in both tasks is the \textit{AssertionError}, which indicates requirements are not fully satisfied in most failed cases. It is worth noting that failed adaptations include more \textit{NameErrors}, \textit{TypeErrors} and \textit{AttributeErrors}, accounting for 47.0\% of all adaptation errors. Specifically, \textit{NameErrors} and \textit{AttributeErrors} are caused by using undefined identifiers in the context. \textit{TypeErrors} contain illegal operands, access violations and method calls inconsistent with their signatures, due to the misapplication of elements in the context. The high proportion of these context-related errors highlights that adaptation exhibits less awareness of the target class context compared to generation.

Through test analysis, we obtain 202 defects with 712 instances from GPT-3.5 adaptations across 200 methods. Their types are illustrated in Table~\ref{tab:undesirable}.
In general, each adaptation case contains an average of 1.01 defects and 3.56 instances. Among 712 defect instances, the \textit{Overlooked} category constitutes 74\%, with 527 instances where GPT-3.5 overlooked pre-existing defects in the given snippet. \textit{Invalid Adapted} defects account for 18\%, which indicates GPT-3.5 could not always resolve identified defects completely during the adaptation. It is worth noting that there are also 54 defect instances categorized as \textit{Unexpected Adapted}, e.g., GPT-3.5 wrongly transformed a static method to a non-static one. 

\begin{table}
    \caption{Identified Defects and Their Types Regarding the Origins}
    \scriptsize
    \renewcommand{\arraystretch}{1.05}
    \begin{center}
        \begin{tabular}{|c|c|c|c|c|c|}
            \hline
            \multirow{2}{*}{} & \multirow{2}{*}{\textbf{\# Defect}} & \multirow{2}{*}{\textbf{\# Defect Inst.}} & \multicolumn{3}{c|}{\textbf{Defect Type}} \\
            \cline{4-6}
            & & & INV & OVE & UNE \\
            \hline
            \textbf{Avg.} & 1.01 & 3.56 & 0.66 & 2.66 & 0.27 \\
            \hline
            \textbf{Total} & 202 & 712 & 131 (18\%) & 527 (74\%) & 54 (8\%) \\
            \hline
            \multicolumn{6}{l}{\makecell[l]{$^*$\textit{OVE}, \textit{INV} and \textit{UNE} denote \emph{Overlooked}, \textit{Invalid Adapted}, and \textit{Unexpected}\\ \textit{Adapted}, respectively.}}
        \end{tabular}
    \end{center}
    \label{tab:undesirable}
\end{table}

\begin{figure}[h]
    \centering
    \includegraphics[width=\linewidth]{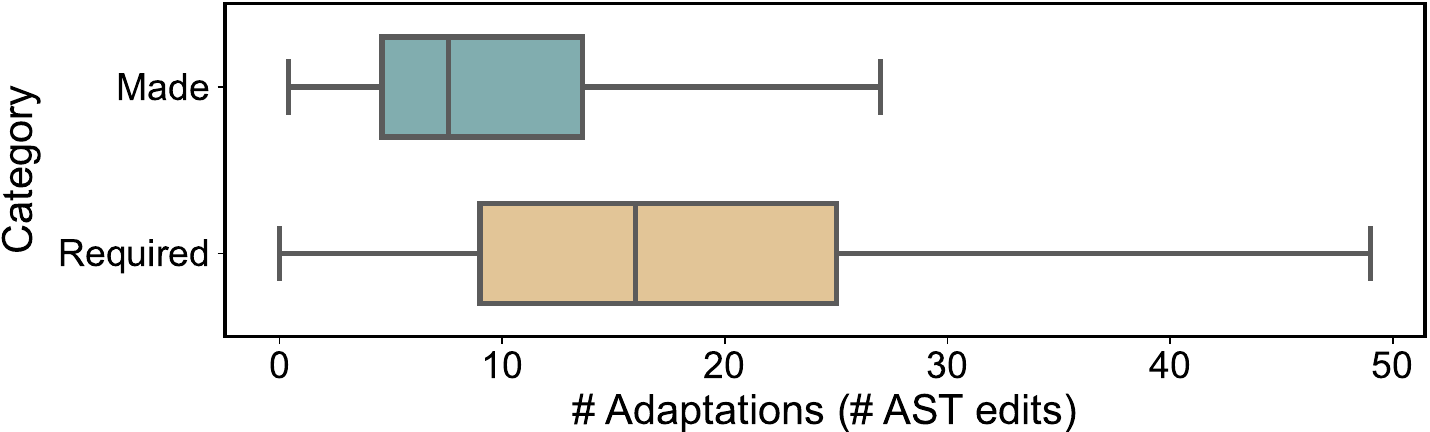}
    \caption{Comparison between the number of adaptations GPT-3.5 actually made and the required adaptation size.}
    \label{fig:boxplot}
\end{figure}

The above result indicates that GPT-3.5 is less capable of addressing the pre-existing defects in retrieved snippets that do not conform to requirements. Furthermore, we compare the number of adaptations done by GPT-3.5 with the required adaptation size, i.e., the number of AST edits needed to obtain the canonical solution, by Mann-Whitney U Test. As illustrated in Fig.~\ref{fig:boxplot}, the number of adaptations done by GPT-3.5 is significant smaller than required adaptations ($p<0.001$, Mann-Whitney U test). It only makes about half of the required adaptations on average. The result implies that GPT-3.5 is ``lazy'' and tend to maintain the original implementation in the given snippet by only making small-scale adaptations, leaving a number of overlooked defects. Therefore, the primary challenge when adapting code with LLMs lies in effectively promoting them in identifying potential defects in previously memorized code snippets.

\begin{center}
    \begin{tcolorbox}[colback=lgray,colframe=black,width=\linewidth,arc=1mm,boxrule=1pt,top=2pt,bottom=2pt,left=3pt,right=3pt]
        \textbf{Finding 2:} Among 200 selected cases, GPT-3.5's adaptations made more context-related errors than generation, indicating its context unawareness in adaptation. The distribution of defect types and the actual adaptation size highlight its laziness on the adaptation task.
    \end{tcolorbox}
\end{center}

\subsection{RQ3: Root Causes of LLMs' Failures}
\label{sec:causes}
Though thematic analysis, we further summarize the root causes of our identified defects. As shown in Table~\ref{tab:causes}, three major categories and nine subcategories are obtained.

\begin{table}
    \caption{Causes For Failures in LLMs' Adaptations}
    \scriptsize
    \renewcommand{\arraystretch}{1.05}
    \begin{center}
        \begin{tabular}{|c|l|r|}
            \hline
            \textbf{Category} & \textbf{Subcategory} & \textbf{\# Defect} \\
            \hline
            \multirow{2}{*}{Unclear Requirement} 
            & Ambiguous Literal & 11 \\
            \cline{2-3}
            & Unspecific Instruction & 2 \\
            \hline
            \multirow{3}{*}{Requirement Misalignment}
            & Method Signature & 40 \\
            \cline{2-3}
            & Operational Logic & 76 \\
            \cline{2-3}
            & Error/Edge Case Handling & 28 \\
            \hline
            \multirow{4}{*}{Context Misapplication}
            & Field Misapplication & 13 \\
            \cline{2-3}
            & Method Misapplication & 11 \\
            \cline{2-3}
            & Environment-related Misapplication & 17 \\
            \cline{2-3}
            & Internal Context Misapplication & 4 \\
            \hline
        \end{tabular}
    \end{center}
    \label{tab:causes}
\end{table}

\textbf{Unclear Requirement} refers to the adaptation failures where developers describe their requirements vaguely or ambiguously in the prompt. LLMs can hardly identify the potential issues in the retrieved snippet and perform intended adaptations without a specific and accurate clarification. Under this category, we identify two subcategories.

\textit{Ambiguous Literal} (11): Developers' requirement may include key concepts with multiple interpretations. Limited by the training data, LLMs are prone to be misled by these ambiguous expressions and generate erroneous adaptations. For instance, the requirement of the \textit{calculate\_sector\_area} method (ClassEval\_5) specifying a parameter \textit{angle} as "\textit{angle of sector}" without indicating whether it is measured in degrees or radians. Likewise, the \textit{FidelityPromo} method (ClassEval\_33) requires LLMs to ``\textit{calculate the discount}" without indicating whether the returned discount is a percentage, a rate or a numerical value. LLMs failed in both cases as they misinterpreted the above concepts. This subcategory indicates LLMs are sensitive to certain literals, thus ambiguous ones may prevent LLMs from correctly adapting the snippet.

\textit{Unspecific Instruction} (2): Developers may use overly concise language to describe their requirement, making it difficult for LLMs to determine the specific details of the task or the desired outcome. This issue is more serious in highly customized and complex task. For instance, the \textit{filter} method (ClassEval\_0) describes its requirement as ``filter the incoming request based on certain rules and conditions." It is hard for LLMs to make adaptations without the specific constraints.

\textbf{Requirement Misalignment} refers to LLMs' failures when aligning the function in the retrieved snippet to the requirement during the adaptation. Different from the \textit{Unclear Requirement} category, developers explain the instructions clearly in the requirement part. This pitfall highlights LLMs' preferences to overgeneralization instead of being restricted.

\textit{Method Signature} (40): This subcategory describes the misalignment of the signature of adapted method. Although we provide the correct and complete signature through the instructions, LLM's adapted method may mismatch the provided signature. The issues include changing a static method to a non-static one, adapting the method with a new name, wrongly adapting the parameter types or numbers, returning a wrong type of value, etc. It indicates that LLMs' inability to subject to the basic constraints for code.

\begin{figure}[H]
    \centering
    \includegraphics[width=\linewidth]{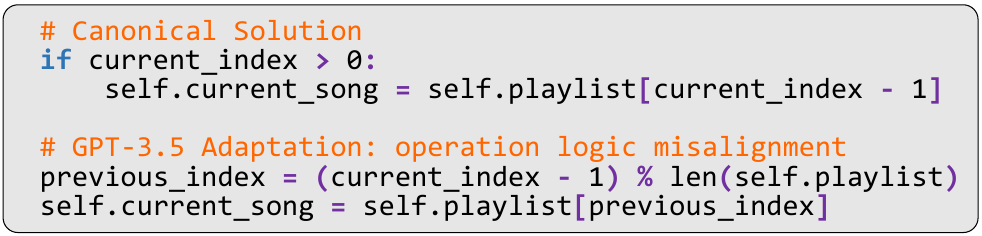}
    \caption{A Failed Adaptation Example for Misalignment in \textit{Operational Logic}.}
    \label{fig:code-2}
\end{figure}

\textit{Operational Logic} (76): This subcategory refers to LLMs' failures to infer and adapt the specific operations the snippet performs, including algorithms, conditionals, and how the code processes the inputs or achieves the desired outputs. Hence, the mismatch of the adapted logic and the requirement leads to failed tests. For instance, the \textit{previous\_song} method (ClassEval\_61) requires the LLM to ``\textit{switch to the previous song in the playlist}" and ``\textit{return False if there was no previous song.}" The LLM retained the looping logic through the playlist, as shown in Fig.~\ref{fig:code-2}. Additionally, LLMs may autonomously apply string case-insensitivity (i.e., ``\textit{lower()}'') or round off numerical computations (i.e., ``\textit{round()}'') during the adaptation. Above cases indicate LLMs' reasoning relies heavily on existing knowledge and statistically frequent solutions, even when they conflict with proposed new requirements.

\textit{Error Handling and Edge Cases} (28): This cause refers to LLMs' inability to handle unexpected inputs and edge cases correctly, focusing on the robustness aspect of their adapted code. LLMs may ignore handling, mishandling or overly handling the special case, which caused the adapted method cannot pass all tests. For example, in the method \textit{calculate\_annulus\_area}, the LLM mishandled the special case when the inner radius of the sector is equals to the outer radius. It returned an error string instead of zero.

\textbf{Context Misapplication} refers to LLMs' misuse of the context knowledge, i.e., data, code, internal representations. We further divided it into four subclasses, including field misapplication, method misapplication, environment-related misapplication and internal context misapplication. 

\begin{figure}[!htbp]
    \centering
    \includegraphics[width=\linewidth]{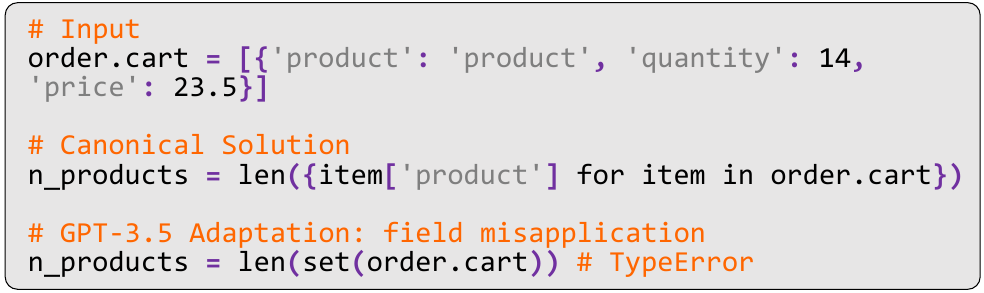}
    \caption{A Failed Adaptation Example for \textit{Field Misapplication}.}
    \label{fig:code-4}
\end{figure}

\textit{Field Misapplication} (13): This refers to LLMs' misinterpretation and misuse of fields. As shown in Fig.~\ref{fig:code-4}, the \textit{cart} field is a list of dictionaries representing different products. The LLM misinterpreted the semantics of the field and converted the list to a set, leading to \textit{TypeError} because the \textit{dict} type is unhashable in Python.

\textit{Method Misapplication} (11): Similarly, method misapplication indicates LLMs' method misuse, e.g., invoking wrong methods, invoking them with wrong parameters, misinterpreting the return value, etc. For instance, in the \textit{solve} method (ClassEval\_35), the LLM mistook the return value of \textit{get\_possible\_moves} method as the next state rather than the move for next state calculation, leading to a \textit{TypeError}.

\begin{figure}[H]
    \centering
    \includegraphics[width=\linewidth]{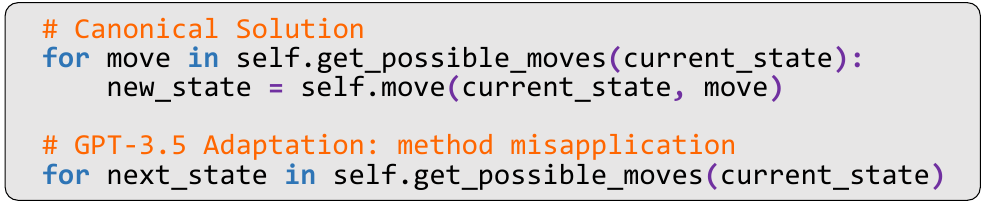}
    \caption{A Failed Adaptation Example for \textit{Method Misapplication}.}
    \label{fig:code-5}
\end{figure}

\textit{Environment-related Misapplication} (17): This refers to LLMs' misinterpretation of the environment, i.e., the scope/boundary of the context and its available resources. For instance, LLMs may use unimported packages, non-existent methods or deprecated APIs. Without the precise perception of the environment, LLMs could also omit the package name, leading to a series of \textit{NameErrors}. 

\begin{figure}[H]
    \centering
    \includegraphics[width=\linewidth]{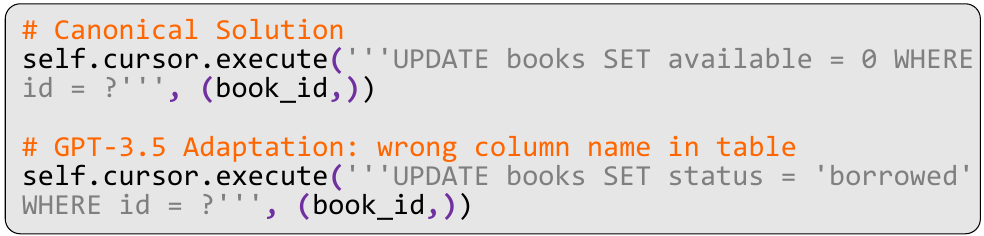}
    \caption{A Failed Adaptation Example for \textit{Internal Context Misapplication}.}
    \label{fig:code-7}
\end{figure}

\textit{Internal Context Misapplication} (4): This refers to LLMs' inability to comprehend and operate the internal context, e.g., tables in the database. As illustrated in Fig.~\ref{fig:code-7}, the LLM failed to identify the schema of the books table through its creation and wrongly applied an \textit{UPDATE} operation to the table and caused \textit{sqlite3.OperationalError}. It raises the concerns for the implicit context when using LLMs to make adaptations in certain domain specific scenarios.

\begin{center}
    \begin{tcolorbox}[colback=lgray,colframe=black,width=\linewidth,arc=1mm,boxrule=1pt,top=2pt,bottom=2pt,left=3pt,right=3pt]
        \textbf{Finding 3:} The main causes for GPT-3.5's adaptation failures are Requirement Misalignment and Context Misapplication, highlighting LLMs' nature of overgeneralization and inattention to constraints. 
        Besides, unclear requirements may also influence their adaptation performance when necessary information are missing.
    \end{tcolorbox}
\end{center}

\section{Our Prompting Approach}
\label{sec3:approach}

Our empirical results demonstrate that the adaptation performance of LLMs is sub-optimal when using current instruction-based prompt (denoted as \textit{Initial Prompt}). Therefore, to address the current pitfalls of LLMs, we propose an interactive prompting approach to utilizing LLMs in code snippet adaptation. We first enhance our prompt by enriching the context and decomposing the task. The resulting prompt is denoted as \textit{Enhanced Prompt}. Additionally, we integrate an interactive workflow to the prompt-based conversation through a ``\textit{flipped interaction}'' process to utilize LLMs' reflection ability. It is implemented with two schemes, one through a human-LLM collaboration and the other through a multi-agent collaboration. The resulting prompts are denoted as \textit{Human-LLM Prompt} and \textit{Multi-Agent Prompt}. The detailed design is described in the following.

\subsection{Prompt Enhancements}

\textbf{Enrich the Context.} In real adaptation scenarios, developers may possess complete method code in their context rather than merely a class skeleton. Therefore, we first consider expanding the class context in the \textit{Initial Prompt} to alleviate the \textit{Context Misapplication} issue. Specifically, we provide docstrings, input-output descriptions, and complete implementations for all the methods independent from the target. This information suggests the function and correct usage of contextual elements. Furthermore, we extract and specify dependencies of the target method to alleviate the \textit{Requirement Misalignment} issue. It pushes LLMs to adapt the misaligned requirements with explicit dependency constraints. We implement this by adding the following statements to the end of the instruction body: ``\textit{It should be implemented using libraries: [Packages], fields: [Fields], methods: [Methods].}'' or ``\textit{It should be implemented without using any external libraries, member variables, or methods.}''

\begin{figure*}[h]
    \centering
    \subfigure[Initial \& Enhanced Prompt]{
        \includegraphics[height=2.05in]{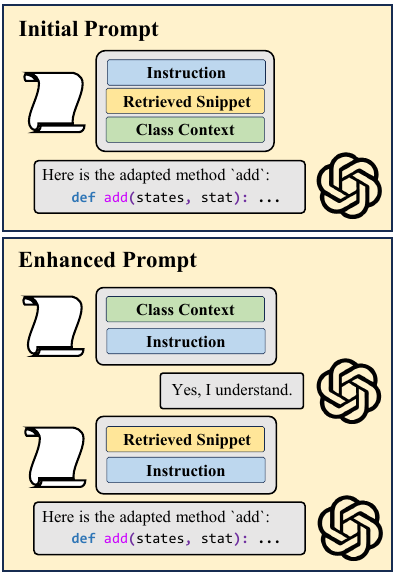}
        \label{fig:init_enha}
    }
    \subfigure[Human-LLM Prompt]{
        \includegraphics[height=2.05in]{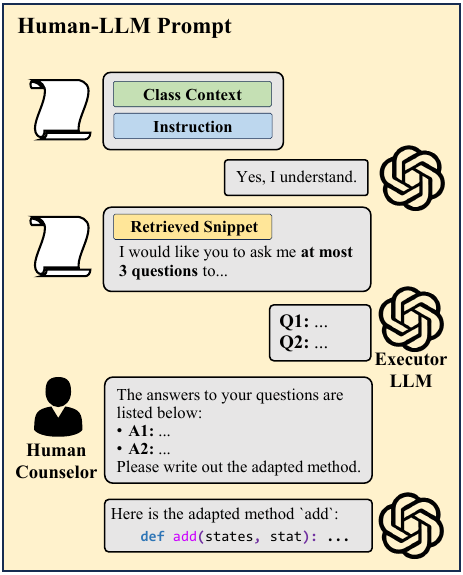}
        \label{fig:human}
    }
    \subfigure[MAC Prompt]{
        \includegraphics[height=2.05in]{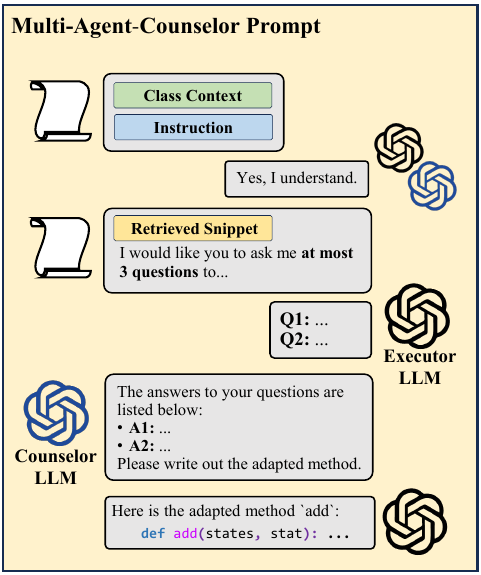}
        \label{fig:with_coun}
    }
    \subfigure[MAE Prompt (Baseline)]{
        \includegraphics[height=2.05in]{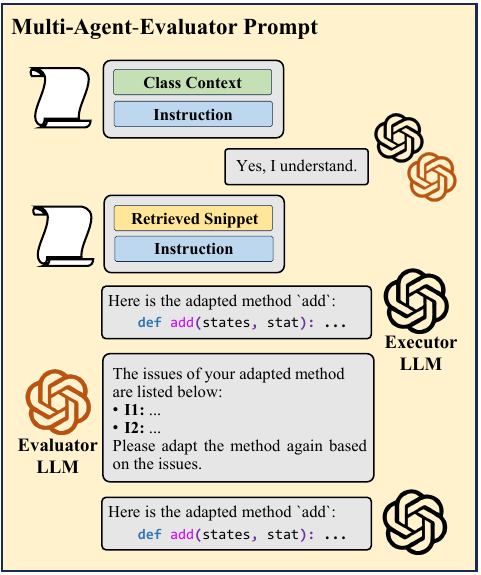}
        \label{fig:with_eval}
    }
    \caption{Illustrations of our proposed prompting approaches.}
    \label{fig:new-prompts}
\end{figure*}

\textbf{Decompose the Task.} The adaptation prompt encompasses requirements, the reused snippet, and the class context. Its overwhelming information leads to LLMs' misinterpretation of the requirements and the context. We consider decomposing the adaptation task into context understanding and adaptation prediction process to further resolve the \textit{Requirement Misalignment} and \textit{Context Misapplication} issues. Specifically, we re-organize our prompt design as a multi-turn conversation, as described in the Fig.~\ref{fig:init_enha}. In the first turn, we only address the context understanding task by simply providing the class context. The followed instruction makes LLMs understand the context but do not generate explanations to reduce costs. In the second turn, we require LLMs to predict adaptations. Splitting the lengthy prompt to shorter ones may alleviate LLMs' burden of handling extremely long context and make them solve the task step by step. Remembering the target class context before adaptation could also refresh their pre-existing knowledge, hence addressing their overgeneralization nature.

\subsection{Interactive Workflow Integration}

Our empirical findings suggest it is challenging for LLMs to adapt code snippets accurately in a single iteration, particularly with \textit{Unclear Requirement}. Drawing inspiration from the communication process between users and developers in software requirement elicitation, we introduce a ``flipped interaction'' process~\cite{White2023a} into the prompting process. Its core idea is to utilize the reflection ability of LLMs. Specifically, we require LLMs to assess the sufficiency and clarity of existing information for the adaptation task. If LLMs are aware of any missed information or any confusion, they could ask questions for more information in the subsequent conversation. This process encourages LLMs to comprehend the adaptation task actively and to refine their memories with information they concern about. To answer the raised questions, we implement this interaction with the following two schemes.

\textbf{Human-LLM Interaction.} The first scheme, \textit{Human-LLM Prompt}, requires developers to provide feedback for LLMs' questions, as shown in Fig.~\ref{fig:human}. To enable a fair comparison with automated prompting approaches, we provide the \textit{Enhanced Prompt} for our participants to understand the requirement, snippet and the target context. Similar to the automated prompting process, our participants should repeatedly interact with LLMs 5 times on each adaptation task to eliminate randomness. To reduce the bias introduced by human repetition, we required participants to provide consistent answers to the same or similar questions asked by LLMs during the experiment. Considering the communication overhead, we conduct the human-LLM interaction on the 200 selected cases with the best GPT-3.5 model, which costs 96 human-hours. All participants have over five years of Python programming experience, which is qualified for the interaction.

\textbf{Multi-Agent Interaction.} Although experienced human developers can pinpoint the problems for adaptation, it is time-consuming for them to employ diagnoses and interactions in practices. To alleviate the human overhead, our second scheme, \textit{Multi-Agent-Counselor (MAC) Prompt}, enables the interaction based on a multi-agent framework. As illustrated in Fig.~\ref{fig:with_coun}, we introduce another LLM as the counselor. In the first turn, it is also initialized with the context understanding prompt. Then the counselor LLM generates answers based on the retrieved snippets and predefined instructions. Finally, the executor LLM accepts the answers and makes adaptations.

Apart from the ``executor-counselor'' interaction, we implement an ``executor-evaluator'' interaction workflow as a baseline, denoted as \textit{Multi-Agent Evaluator (MAE) Prompt}, leveraging LLMs' ability to assess generated code~\cite{Zhuo2023}. Fig.~\ref{fig:with_eval} illustrates its process. An evaluator LLM, also initialized with the context understanding prompt, reviews the adapted method generated by the executor LLM. Then it generates a list of issues based on its assessment and instructs the executor to regenerated an adaptation to address the issues. All agents above are profiled using role-playing system prompts without further training or fine-tuning.

\textbf{Configurations of the Interactive Workflow.} As we require LLMs to assess the sufficiency of information in the interactive workflow of the \textit{Human-LLM} and \textit{MAC Prompt}, it is uncertain when LLMs will terminate the conversation. Therefore, we conducted a preliminary experiment to determine two relevant configurations, the number of iterations and the number of questions. To this end, we randomly picked 20 methods from outside of our 200 selected cases and conducted the interaction with GPT-3.5, where the question number was not restricted. In all 20 cases, GPT-3.5 generated code right after an iteration in which all the answers were provided. They could ask up to 8 questions with the average number is 3.2. Therefore, as illustrated in Fig.~\ref{fig:new-prompts}, our prompt limits LLMs to ask at most 3 questions and extracts the code block after one iteration. If no code block is provided, the interaction will still be terminated and return an empty string.

\section{Evaluation}
\label{sec4:evaluation}

\begin{table}
    \caption{Adaptation Performance with Enhanced Prompt}
    \renewcommand{\arraystretch}{1.05}
    \begin{center}
        \begin{tabular}{|l|c|c|c|}
            \hline
            \textbf{Model} & \textbf{pass@1} & \textbf{pass@5} & \textbf{CodeBLEU} \\
            \hline
            GPT-3.5 & 52.34 & 60.98 & 47.05 \\
            GPT-3.5-Enhanced & \textbf{67.76} & \textbf{74.39} & \textbf{60.16} \\
            \hline
            \textbf{w/o} docstring \& code & 59.22 & 66.34 & 52.44 \\
            % w/o code context & 62.68 & 70.00 & 53.67 \\
            \textbf{w/o} dependency & 66.63 & 72.44 & 58.50 \\
            \textbf{w/o} decomposition & 62.15 & 70.98 & 56.45 \\
            \hline
        \end{tabular}
    \end{center}
    \label{tab:enhanced}
\end{table}

In this section, we evaluate the effectiveness of our prompt enhancements and interactive workflow on the adaptation task.

\subsection{Effectiveness of the \textit{Enhanced Prompt}} 
Table~\ref{tab:enhanced} presents the adaptation performance of prompting GPT-3.5 with the \textit{Enhanced Prompt}. It achieves a pass@1 score of 67.76\% and a pass@5 score of 74.39\%, which represents an improvement of 29.46\% and 21.99\% compared to the \textit{Initial Prompt}. Furthermore, its CodeBLEU increases from 47.05 to 60.16, marking a 27.86\% relative improvement. The result indicates that \textit{Enhanced Prompt} substantially enhances the performance of LLMs in the adaptation task.
% ablation study results
We also conduct an ablation study to examine the effects of three prompt enhancements within the \textit{Enhanced Prompt}. Among them, enlarging the code context offers the most significant assistance to GPT-3.5. It results in an 8.54\% gain in pass@1 and an 8.05\% and pass@5. The decomposition strategy is also effective, marking a 5.61\% improvement in pass@1. The dependency information exhibits a slight improvement in LLMs' adaptation performance.

\subsection{Effectiveness of the \textit{Human-LLM Prompt}}

\begin{table}
    \caption{Adaptation Performance with Interactive Workflow on 200 Selected Cases}
    \renewcommand{\arraystretch}{1.05}
    \begin{center}
        \begin{tabular}{|l|c|c|c|}
            \hline
            \textbf{Model} & \textbf{pass@1} & \textbf{pass@5} & \textbf{CodeBLEU} \\
            \hline
            GPT-3.5 & 52.70 & 61.00 & 48.02 \\
            GPT-3.5-Enhanced & 71.20 & 78.00 & \textbf{61.61}  \\
            GPT-3.5-Human-LLM & \textbf{74.50} & \textbf{87.00} & 60.53 \\
            \hline
            GPT-3.5-MAC & 71.80 & 82.50 & 60.45 \\
            GPT-3.5-MAE (baseline) & 65.20 & 80.00 & 57.32 \\
            % GPT-3.5-Both & 63.70 & 82.00 & 56.36 \\
            \hline
        \end{tabular}
    \end{center}
    \label{tab:interactive}
\end{table}

The adaptation performance of the \textit{Human-LLM Prompt} on our 200 selected cases is shown in Table~\ref{tab:interactive}. It achieves the best performance among all prompt design. Guided with the human-written response, GPT-3.5 obtain 74.50 in pass@1 and 87.00 in pass@5, respectively, which outperforms the \textit{Initial Prompt} by 41.4\% and 42.6\%. However, its CodeBLEU value is slightly lower than the \textit{Enhanced Prompt}. The reason is that the executor tends to make adaptations corresponding to various questions that the counselor has asked, leading to differences in the code structure. The decrease in this heuristic metric does not harm the accuracy of adaptation results.

\begin{table}
    \caption{Mitigation of Defects in Adaptation}
    \renewcommand{\arraystretch}{1.05}
    \scriptsize
    \begin{center}
        \begin{tabular}{|l|c|c|}
            \hline
            \textbf{Cause} & \textbf{GPT-3.5} & \textbf{GPT-3.5-Human-LLM} \\
            \hline
            Ambiguous Literal & 11 & 3 (-72.7\%) \\
            Unspecific Instruction & 2 & 2 (-0.0\%) \\
            \hline
            Method Signature & 40 & 6 (-85.0\%) \\
            Operational Logic & 76 & 15 (-80.3\%) \\
            Error/Edge Case Handling & 28 & \ \:7 (-75.0\%) \\
            \hline
            Field Misapplication & 13 & 2 (-84.6\%) \\
            Method Misapplication & 11 & 4 (-63.6\%) \\ 
            Environment-related Misapplication & 17 & 4 (-76.5\%) \\
            Internal Context Misapplication & 4 & 0 (-100\%) \\
            \hline
            \textbf{Total} & \textbf{202} & \textbf{43 (-79.2\%)} \\
            \hline
        \end{tabular}
    \end{center}
    \label{tab:mitigation}
\end{table}

We further inspect how our identified defects are mitigated by our prompt enhancements and interactive workflow. As shown in Table~\ref{tab:mitigation}, the \textit{Human-LLM Prompt} solves 159 out of 202 defects, which accounts for 78.7\%. Our approach is effective in resolving defects across most categories, particularly for \textit{Method Signature}, \textit{Field Misapplication} and \textit{Internal Context Misapplication}. However, two defects caused by \textit{Unspecific Instruction} are still unresolved even with human feedback. The reason may be that the task is highly customized and contains a wealth of contextual knowledge, which could be hard to explain in the interaction.
As for the communication overhead, totally 1,416, 571, and 45 selective, close-ended and open-ended questions are asked by the LLM respectively. For each case, developers spend over 5 minutes ($96/1000=5.76$) on average to review the context and synthesize an effective instruction, which introduces immeasurable overhead in real scenarios. Therefore, we investigate the possibility of enabling automatic interaction based on the Multi-Agent workflow.

\subsection{Effectiveness of the \textit{Multi-Agent Prompt}}

As illustrated in Table~\ref{tab:interactive}, the \textit{MAC Prompt} achieves 71.80, 82.50 and 60.45 in pass@1, pass@5 and CodeBLEU. It could further improve the pass@5 score by 4.50 compared to the \textit{Enhanced Prompt}. For the \textit{MAE Prompt} baseline, it is inferior to the \textit{MAC Prompt} on all metrics and obtains an extremely low pass@1, even lower than the \textit{Enhanced Prompt}. It indicates that the Evaluator tends to provide inaccurate feedback to the Executor, leading to extra erroneous adaptations. Note that although the \textit{Human-LLM Prompt} still hits the ceiling with the best adaptation performance, its human-in-the-loop interaction may limit the practical use of this approach. The \textit{MAC Prompt} could be an excellent trade-off between performance and overhead, as supported by our experimental results.

\begin{table}
    \caption{Adaptation Performance with Multi-Agent-Counselor Prompt on Instruction-Tuned LLMs}
    \renewcommand{\arraystretch}{1.05}
    \begin{center}
        \begin{tabular}{|l|c|c|c|}
            \hline
            \textbf{Model} & \textbf{pass@1} & \textbf{pass@5} & \textbf{CodeBLEU} \\
            \hline
            CodeLlama-34B & 32.78 & 49.51 & 37.56 \\
            CodeLlama-34B-Enhanced & 41.02 & 65.37 & 40.83 \\
            CodeLlama-34B-MAC & \textbf{45.95} & \textbf{67.32} & \textbf{45.58} \\
            \hline
            Llama-3-8B & 42.24 & 58.29 & 45.34 \\
            Llama-3-8B-Enhanced & \textbf{57.51} & 68.78 & \textbf{53.70} \\
            Llama-3-8B-MAC & 54.05 & \textbf{72.20} & 51.34 \\
            \hline
        \end{tabular}
    \end{center}
    \label{tab:generalization}
\end{table}

To further explore the application of the \textit{MAC Prompt}, we evaluate its effectiveness with our instruction-tuned LLMs on the whole \textit{ClassEval} dataset. We choose LLMs with the best performing settings, i.e., CodeLlama-34B-temp0.6 and Llama-3-8B-temp0.6.
As shown in Table~\ref{tab:generalization}, our Enhanced Prompt also brings a clear benefit for instruction-tuned LLMs. Our proposed \textit{MAC Prompt} could further improve the pass@1, pass@5 and CodeBLEU score of CodeLlama-34B by 4.93, 1.95 and 4.75, respectively. For Llama-3-8B, the \textit{MAC Prompt} results in a 3.42 increase in pass@5 but with a slightly decrease in the other two metrics. The result highlights the generalization ability of our prompting approach. Both our prompt enhancements and interaction workflow are effective for the adaptation task regardless of the model.

\begin{center}
    \begin{tcolorbox}[colback=lgray,colframe=black,width=\linewidth,arc=1mm,boxrule=1pt,top=2pt,bottom=2pt,left=3pt,right=3pt]
        \textbf{Finding 4:} Our proposed prompt enhancements and interactive workflow greatly improve LLMs' adaptation performance. Human-LLM interaction achieves the ideal performance and resolves most identified defects, while our MAC interaction can fully automate this process with a similar performance and no human intervention.
        
    \end{tcolorbox}
\end{center}

\section{Related Work}
\label{sec5:rw}

\subsection{Code Snippet Adaptation}
Prior studies on adaptation mainly focus on improving the efficiency. Cottrell et al.~\cite{Cottrell2008} present Jigsaw to integrate snippets into developers' code, which reduces their efforts on simple adaptations such as renaming. Wightman et al.~\cite{Wightman2012} develop a lightweight Eclipse plug-in, SnipMatch. It supports the search and integration of code templates in the code context. Zhang et al.~\cite{Zhang2019} develop a Chrome plugin, ExampleStack, to generate templates from developers' historical modifications on code snippets. Reid et al.~\cite{Reid2020} propose NLP2TestableCode to reduce the compilation errors during adaptation and generate tests based on input/output types in the context. Terragni et al.~\cite{Terragni2021} present APIZATOR to transform code snippets on Stack Overflow to well-formed methods. However, there is not a general-purpose predictive approach for automated adaptation. The emergence of LLMs provides opportunities for this task because of their strong natural language processing ability. This paper explores their adaptation capabilities to provide insights for this research field. 

\subsection{Prompt Engineering for Software Engineering Tasks}
Modern LLMs have demonstrated their capabilities in various software tasks~\cite{Tian2023,Sridhara2023}, including code generation, program repair~\cite{Sobania2023,Xia2023,lin2024one} and code maintenance~\cite{Kabir2023,Guo2024,lin2023cct5,wang2024divide}.
Enhancing LLMs' performance in these domains often involves prompt engineering. This includes handcrafted prompt design, the Chain-of-Thought (CoT) strategy~\cite{Wei2023}, and constructing effective demonstrations~\cite{Min2022,Nashid2023,Gao2023}.

Specifically, Liu et al.~\cite{Liu2023} improve code generation prompts using the CoT strategy with multi-step optimization. Cao et al.~\cite{Cao2023} explore various prompt templates for deep learning-based program repair. 
% Ahmad et al.~\cite{Ahmad2023} provide more semantic information for the few-shot prompts in the code summarization task.
Rodriguez et al.~\cite{Rodriguez2023} boost LLMs' performance in software traceability by guiding them to perform intermediate reasoning. Gao et al.~\cite{Gao2023} focus on constructing effective demonstrations for code intelligence tasks. These strategies have shown significant impact on their respective tasks, but their applicability to other tasks, i.e., code snippet adaptation, needs further exploration.

Furthermore, prior work has presented reusable prompting patterns for LLMs in software engineering~\cite{White2023a,White2023b}. White et al.~\cite{White2023a} introduce a structured prompt framework with general patterns, including the flipped interaction pattern, which enables LLMs to gather information actively. This pattern is further extended in our work to support requirement refinement in code snippet adaptation.

\section{Threats to Validity}
\label{sec6:threats}

\textbf{Internal Validity.} 
1) The inherent randomness in LLM outputs, particularly in GPT-3.5, can lead to various adaptations. We mitigated this by requesting results five times from the LLMs and analyzing them using pass@1, pass@5, and CodeBLEU metrics, aiming for a more reliable assessment of their performance;
2) Our study only considers using three popular LLMs due to the expense. Involving more and newer models could provide more insights and potentially improve the performance;
3) This paper focuses on utilizing LLMs for software reuse. This paradigm could be changed when generative AIs are powerful enough to build any software satisfying developers' needs from scratch.

\textbf{External Validity.} 
1) In the snippet retrieval phase, we use LLM-generated code instead of real snippets from open-source communities, which may not fully represent real-world adaptation scenarios. To better understand LLMs' adaptations, future studies could build a comprehensive dataset based on developers adaptation practices, e.g., adaptations from Stack Overflow, to further validate the generalization of our findings;
2) In this paper, we construct the adaptation cases based on the handcrafted \textit{ClassEval} dataset with only 410 Python methods. However, they may not fully capture the complexity and diversity of real-world adaptation scenarios. To support adaptations in a broad context, future research could integrate an effective context retrieval module to our approach in real application scenarios;
3) Our empirical study and experimental evaluation only focus on three representative LLMs. Different LLMs could have different distributions of their adaptation issues and root causes. It is also worth investigating whether our approach is effective for other series of LLMs. However, our identification of LLMs' failure reasons is actually independent of the model and our inspection on adequate samples could reveal useful findings to LLMs' adaptations. 

\textbf{Construct Validity.} 
1) The human-designed adaptation prompts in our study might not represent the optimal performance for adaptation tasks. There is potential for alternative, possibly more effective, prompts, including those generated automatically by LLMs as suggested in other studies (e.g., Liu et al.~\cite{Liu2023}). This highlights a need for further exploration of prompt design methodologies;
2) The subjective nature of our manual analysis process could introduce biases. To mitigate this, we conduct our manual analysis in a rigorous manner. Our reported Cohen's Kappa indicates a high agreement between annotators and hence mitigate this threat to an extent. Nevertheless, the potential for subjective interpretation remains a factor to be considered in the evaluation of our results.

\section{Conclusion}
\label{sec7:conclusion}
In this paper, we first conduct an empirical study to explore LLMs' capabilities in code snippet adaptation. Due to their sub-optimal performance compared to code generation, we further investigate the problems and causes by analyzing the adapted snippet along with their test results. Based on our empirical findings, we propose an interactive prompting approach to mitigating current problems. It enables the reflection ability of LLMs to improve their context awareness. Our evaluation result shows the effectiveness of our approach. Human-LLM interaction could achieve the best performance but introduce human efforts. We propose a multi-agent interaction that could achieve comparable performance as a trade-off. Its generalization ability is also validated. Our research findings provide insights for LLMs' current limitations and highlight the effectiveness of interactive prompting. We believe that our approach could facilitate the practical application of LLMs in real-world snippet reuse and adaptation.

\section*{Acknowledgments}
We gratefully acknowledge the support from the National Natural Science Foundation of China (Grant No.62302515, No.62172426, and No.62332005), the National Key Research and Development Program of China (Grant No.2023YFB4503802), and the National University of Defense Technology Research Project (Grant No.ZK21-13).

\bibliographystyle{ieeetr}
\bibliography{ref}

\begin{thebibliography}{10}

\bibitem{Yang2017}
D.~Yang, P.~Martins, V.~Saini, and C.~Lopes, ``Stack {Overflow} in {Github}:
  {Any} {Snippets} {There}?,'' in {\em 2017 {IEEE}/{ACM} 14th {International}
  {Conference} on {Mining} {Software} {Repositories} ({MSR})}, pp.~280--290,
  2017.

\bibitem{Baltes2019}
S.~Baltes and S.~Diehl, ``Usage and attribution of {Stack} {Overflow} code
  snippets in {GitHub} projects,'' {\em Empirical Software Engineering},
  vol.~24, pp.~1259--1295, June 2019.

\bibitem{Manes2021}
S.~S. Manes and O.~Baysal, ``Studying the {Change} {Histories} of {Stack}
  {Overflow} and {GitHub} {Snippets},'' in {\em 2021 {IEEE}/{ACM} 18th
  {International} {Conference} on {Mining} {Software} {Repositories} ({MSR})},
  (Madrid, Spain), pp.~283--294, IEEE, May 2021.

\bibitem{Huang2022}
Y.~Huang, F.~Xu, H.~Zhou, X.~Chen, X.~Zhou, and T.~Wang, ``Towards {Exploring}
  the {Code} {Reuse} from {Stack} {Overflow} during {Software} {Development},''
  in {\em Proceedings of the 30th {IEEE}/{ACM} {International} {Conference} on
  {Program} {Comprehension}}, {ICPC} '22, (New York, NY, USA), pp.~548--559,
  Association for Computing Machinery, 2022.
\newblock event-place: Virtual Event.

\bibitem{wang2023natural}
S.~Wang, M.~Geng, B.~Lin, Z.~Sun, M.~Wen, Y.~Liu, L.~Li, T.~F. Bissyand{\'e},
  and X.~Mao, ``Natural language to code: How far are we?,'' in {\em
  Proceedings of the 31st ACM Joint European Software Engineering Conference
  and Symposium on the Foundations of Software Engineering}, pp.~375--387,
  2023.

\bibitem{wang2023two}
S.~Wang, B.~Lin, Z.~Sun, M.~Wen, Y.~Liu, Y.~Lei, and X.~Mao, ``Two birds with
  one stone: Boosting code generation and code search via a generative
  adversarial network,'' {\em Proceedings of the ACM on Programming Languages},
  vol.~7, no.~OOPSLA2, pp.~486--515, 2023.

\bibitem{Brandt2009}
J.~Brandt, P.~J. Guo, J.~Lewenstein, M.~Dontcheva, and S.~R. Klemmer, ``Writing
  {Code} to {Prototype}, {Ideate}, and {Discover},'' {\em IEEE Software},
  vol.~26, no.~5, pp.~18--24, 2009.

\bibitem{Yang2016}
D.~Yang, A.~Hussain, and C.~V. Lopes, ``From query to usable code: an analysis
  of stack overflow code snippets,'' in {\em Proceedings of the 13th
  {International} {Conference} on {Mining} {Software} {Repositories}}, (Austin
  Texas), pp.~391--402, ACM, May 2016.

\bibitem{wang2024fusing}
S.~Wang, M.~Geng, B.~Lin, Z.~Sun, M.~Wen, Y.~Liu, L.~Li, T.~F. Bissyand{\'e},
  and X.~Mao, ``Fusing code searchers,'' {\em IEEE Transactions on Software
  Engineering}, 2024.

\bibitem{Zhang2019}
T.~Zhang, D.~Yang, C.~Lopes, and M.~Kim, ``Analyzing and {Supporting}
  {Adaptation} of {Online} {Code} {Examples},'' in {\em Proceedings of the 41st
  {International} {Conference} on {Software} {Engineering}}, {ICSE} '19,
  pp.~316--327, IEEE Press, 2019.
\newblock event-place: Montreal, Quebec, Canada.

\bibitem{Mondal2019}
M.~Mondal, B.~Roy, C.~K. Roy, and K.~A. Schneider, ``Investigating {Context}
  {Adaptation} {Bugs} in {Code} {Clones},'' in {\em 2019 {IEEE} {International}
  {Conference} on {Software} {Maintenance} and {Evolution} ({ICSME})},
  (Cleveland, OH, USA), pp.~157--168, IEEE, Sept. 2019.

\bibitem{Zhang2024}
T.~Zhang, Y.~Lu, Y.~Yu, X.~Mao, Y.~Zhang, and Y.~Zhao, ``How do developers
  adapt code snippets to their contexts? an empirical study of context-based
  code snippet adaptations,'' {\em IEEE Transactions on Software Engineering},
  vol.~50, no.~11, pp.~2712--2731, 2024.

\bibitem{Cottrell2008}
R.~Cottrell, R.~J. Walker, and J.~Denzinger, ``Jigsaw: a tool for the
  small-scale reuse of source code,'' in {\em Companion of the 13th
  international conference on {Software} engineering - {ICSE} {Companion} '08},
  (Leipzig, Germany), p.~933, ACM Press, 2008.

\bibitem{Wightman2012}
D.~Wightman, Z.~Ye, J.~Brandt, and R.~Vertegaal, ``{SnipMatch}: {Using}
  {Source} {Code} {Context} to {Enhance} {Snippet} {Retrieval} and
  {Parameterization},'' pp.~219--228, ACM, 2012.

\bibitem{Reid2020}
B.~Reid, C.~Treude, and M.~Wagner, ``Optimising the fit of stack overflow code
  snippets into existing code,'' in {\em Proceedings of the 2020 {Genetic} and
  {Evolutionary} {Computation} {Conference} {Companion}}, (Cancún Mexico),
  pp.~1945--1953, ACM, July 2020.

\bibitem{Terragni2021}
V.~Terragni and P.~Salza, ``{APIzation}: {Generating} {Reusable} {APIs} from
  {StackOverflow} {Code} {Snippets},'' in {\em 2021 36th {IEEE}/{ACM}
  {International} {Conference} on {Automated} {Software} {Engineering}
  ({ASE})}, (Melbourne, Australia), pp.~542--554, IEEE, Nov. 2021.

\bibitem{OpenAI2023}
OpenAI, ``gpt-3.5-turbo,'' 2023.

\bibitem{Brown2020}
T.~B. Brown, B.~Mann, N.~Ryder, M.~Subbiah, J.~Kaplan, P.~Dhariwal,
  A.~Neelakantan, P.~Shyam, G.~Sastry, A.~Askell, S.~Agarwal, A.~Herbert-Voss,
  G.~Krueger, T.~Henighan, R.~Child, A.~Ramesh, D.~M. Ziegler, J.~Wu,
  C.~Winter, C.~Hesse, M.~Chen, E.~Sigler, M.~Litwin, S.~Gray, B.~Chess,
  J.~Clark, C.~Berner, S.~McCandlish, A.~Radford, I.~Sutskever, and D.~Amodei,
  ``Language {Models} are {Few}-{Shot} {Learners},'' July 2020.
\newblock arXiv:2005.14165 [cs].

\bibitem{Wei2022}
J.~Wei, Y.~Tay, R.~Bommasani, C.~Raffel, B.~Zoph, S.~Borgeaud, D.~Yogatama,
  M.~Bosma, D.~Zhou, D.~Metzler, E.~H. Chi, T.~Hashimoto, O.~Vinyals, P.~Liang,
  J.~Dean, and W.~Fedus, ``Emergent {Abilities} of {Large} {Language}
  {Models},'' Oct. 2022.
\newblock arXiv:2206.07682 [cs].

\bibitem{Gao2023}
S.~Gao, X.-C. Wen, C.~Gao, W.~Wang, H.~Zhang, and M.~R. Lyu, ``What {Makes}
  {Good} {In}-context {Demonstrations} for {Code} {Intelligence} {Tasks} with
  {LLMs}?,'' Aug. 2023.
\newblock arXiv:2304.07575 [cs].

\bibitem{Du2023}
X.~Du, M.~Liu, K.~Wang, H.~Wang, J.~Liu, Y.~Chen, J.~Feng, C.~Sha, X.~Peng, and
  Y.~Lou, ``Evaluating large language models in class-level code generation,''
  in {\em Proceedings of the IEEE/ACM 46th International Conference on Software
  Engineering}, ICSE '24, (New York, NY, USA), Association for Computing
  Machinery, 2024.

\bibitem{geng2024large}
M.~Geng, S.~Wang, D.~Dong, H.~Wang, G.~Li, Z.~Jin, X.~Mao, and X.~Liao, ``Large
  language models are few-shot summarizers: Multi-intent comment generation via
  in-context learning,'' in {\em Proceedings of the 46th IEEE/ACM International
  Conference on Software Engineering}, pp.~1--13, 2024.

\bibitem{Sobania2023}
D.~Sobania, M.~Briesch, C.~Hanna, and J.~Petke, ``An {Analysis} of the
  {Automatic} {Bug} {Fixing} {Performance} of {ChatGPT},'' Jan. 2023.
\newblock arXiv:2301.08653 [cs].

\bibitem{Cao2023}
J.~Cao, M.~Li, M.~Wen, and S.-c. Cheung, ``A study on {Prompt} {Design},
  {Advantages} and {Limitations} of {ChatGPT} for {Deep} {Learning} {Program}
  {Repair},'' Apr. 2023.
\newblock arXiv:2304.08191 [cs].

\bibitem{Xia2023}
C.~S. Xia and L.~Zhang, ``Keep the {Conversation} {Going}: {Fixing} 162 out of
  337 bugs for \$0.42 each using {ChatGPT},'' Apr. 2023.
\newblock arXiv:2304.00385 [cs].

\bibitem{qin2024agentfl}
Y.~Qin, S.~Wang, Y.~Lou, J.~Dong, K.~Wang, X.~Li, and X.~Mao, ``Agentfl:
  Scaling llm-based fault localization to project-level context,'' {\em arXiv
  preprint arXiv:2403.16362}, 2024.

\bibitem{Liu2021}
P.~Liu, W.~Yuan, J.~Fu, Z.~Jiang, H.~Hayashi, and G.~Neubig, ``Pre-train,
  {Prompt}, and {Predict}: {A} {Systematic} {Survey} of {Prompting} {Methods}
  in {Natural} {Language} {Processing},'' July 2021.
\newblock arXiv:2107.13586 [cs].

\bibitem{Reynolds2021}
L.~Reynolds and K.~McDonell, ``Prompt {Programming} for {Large} {Language}
  {Models}: {Beyond} the {Few}-{Shot} {Paradigm},'' Feb. 2021.
\newblock arXiv:2102.07350 [cs].

\bibitem{Rodriguez2023}
A.~D. Rodriguez, K.~R. Dearstyne, and J.~Cleland-Huang, ``Prompts {Matter}:
  {Insights} and {Strategies} for {Prompt} {Engineering} in {Automated}
  {Software} {Traceability},'' July 2023.
\newblock arXiv:2308.00229 [cs].

\bibitem{Liu2022}
V.~Liu and L.~B. Chilton, ``Design {Guidelines} for {Prompt} {Engineering}
  {Text}-to-{Image} {Generative} {Models},'' in {\em {CHI} {Conference} on
  {Human} {Factors} in {Computing} {Systems}}, (New Orleans LA USA), pp.~1--23,
  ACM, Apr. 2022.

\bibitem{Luo2023}
Z.~Luo, C.~Xu, P.~Zhao, Q.~Sun, X.~Geng, W.~Hu, C.~Tao, J.~Ma, Q.~Lin, and
  D.~Jiang, ``{WizardCoder}: {Empowering} {Code} {Large} {Language} {Models}
  with {Evol}-{Instruct},'' June 2023.
\newblock arXiv:2306.08568 [cs].

\bibitem{Zhao2023}
W.~X. Zhao, K.~Zhou, J.~Li, T.~Tang, X.~Wang, Y.~Hou, Y.~Min, B.~Zhang,
  J.~Zhang, Z.~Dong, Y.~Du, C.~Yang, Y.~Chen, Z.~Chen, J.~Jiang, R.~Ren, Y.~Li,
  X.~Tang, Z.~Liu, P.~Liu, J.-Y. Nie, and J.-R. Wen, ``A {Survey} of {Large}
  {Language} {Models},'' June 2023.
\newblock arXiv:2303.18223 [cs].

\bibitem{Cohen1960}
J.~Cohen, ``A coefficient of agreement for nominal scales,'' {\em Educational
  and Psychological Measurement}, vol.~20, pp.~37 -- 46, 1960.

\bibitem{Braun2006}
V.~Braun and V.~Clarke, ``Using thematic analysis in psychology,'' {\em
  Qualitative Research in Psychology}, vol.~3, no.~2, pp.~77--101, 2006.

\bibitem{Berg2017}
B.~L. Berg and H.~Lune, {\em Qualitative research methods for the social
  sciences}.
\newblock Books a la carte, Boston: Pearson, ninth edition~ed., 2017.

\bibitem{Roziere2023}
B.~Rozière, J.~Gehring, F.~Gloeckle, S.~Sootla, I.~Gat, X.~E. Tan, Y.~Adi,
  J.~Liu, T.~Remez, J.~Rapin, A.~Kozhevnikov, I.~Evtimov, J.~Bitton, M.~Bhatt,
  C.~C. Ferrer, A.~Grattafiori, W.~Xiong, A.~Défossez, J.~Copet, F.~Azhar,
  H.~Touvron, L.~Martin, N.~Usunier, T.~Scialom, and G.~Synnaeve, ``Code
  {Llama}: {Open} {Foundation} {Models} for {Code},'' Aug. 2023.
\newblock arXiv:2308.12950 [cs].

\bibitem{Meta2024}
``Meta llama 3.'' \url{https://llama.meta.com/}.
\newblock Accessed: 2024-04-28.

\bibitem{Santu2023}
S.~K.~K. Santu and D.~Feng, ``{TELeR}: {A} {General} {Taxonomy} of {LLM}
  {Prompts} for {Benchmarking} {Complex} {Tasks},'' May 2023.
\newblock arXiv:2305.11430 [cs].

\bibitem{Ren2020}
S.~Ren, D.~Guo, S.~Lu, L.~Zhou, S.~Liu, D.~Tang, N.~Sundaresan, M.~Zhou,
  A.~Blanco, and S.~Ma, ``{CodeBLEU}: a {Method} for {Automatic} {Evaluation}
  of {Code} {Synthesis},'' Sept. 2020.
\newblock arXiv:2009.10297 [cs].

\bibitem{Holtzman2020}
A.~Holtzman, J.~Buys, L.~Du, M.~Forbes, and Y.~Choi, ``The curious case of
  neural text degeneration,'' in {\em International Conference on Learning
  Representations}, 2020.

\bibitem{Mann1946}
H.~Mann and D.~Whitney, ``On a test of whether one of two random variables is
  stochastically larger than the other,'' {\em Annals of Mathematical
  Statistics}, vol.~18, 11 1946.

\bibitem{White2023a}
J.~White, Q.~Fu, S.~Hays, M.~Sandborn, C.~Olea, H.~Gilbert, A.~Elnashar,
  J.~Spencer-Smith, and D.~C. Schmidt, ``A {Prompt} {Pattern} {Catalog} to
  {Enhance} {Prompt} {Engineering} with {ChatGPT},'' Feb. 2023.
\newblock arXiv:2302.11382 [cs].

\bibitem{Zhuo2023}
T.~Y. Zhuo, ``Large {Language} {Models} {Are} {State}-of-the-{Art} {Evaluators}
  of {Code} {Generation},'' Apr. 2023.
\newblock arXiv:2304.14317 [cs].

\bibitem{Tian2023}
H.~Tian, W.~Lu, T.~O. Li, X.~Tang, S.-C. Cheung, J.~Klein, and T.~F.
  Bissyandé, ``Is {ChatGPT} the {Ultimate} {Programming} {Assistant} -- {How}
  far is it?,'' Apr. 2023.
\newblock arXiv:2304.11938 [cs].

\bibitem{Sridhara2023}
G.~Sridhara, R.~H. G., and S.~Mazumdar, ``{ChatGPT}: {A} {Study} on its
  {Utility} for {Ubiquitous} {Software} {Engineering} {Tasks},'' May 2023.
\newblock arXiv:2305.16837 [cs].

\bibitem{lin2024one}
B.~Lin, S.~Wang, M.~Wen, L.~Chen, and X.~Mao, ``One size does not fit all:
  Multi-granularity patch generation for better automated program repair,'' in
  {\em Proceedings of the 33rd ACM SIGSOFT International Symposium on Software
  Testing and Analysis}, pp.~1554--1566, 2024.

\bibitem{Kabir2023}
M.~M.~A. Kabir, S.~A. Hassan, X.~Wang, Y.~Wang, H.~Yu, and N.~Meng, ``An
  empirical study of {ChatGPT}-3.5 on question answering and code
  maintenance,'' Oct. 2023.
\newblock arXiv:2310.02104 [cs].

\bibitem{Guo2024}
Q.~Guo, J.~Cao, X.~Xie, S.~Liu, X.~Li, B.~Chen, and X.~Peng, ``Exploring the
  {Potential} of {ChatGPT} in {Automated} {Code} {Refinement}: {An} {Empirical}
  {Study},'' 2024.

\bibitem{lin2023cct5}
B.~Lin, S.~Wang, Z.~Liu, Y.~Liu, X.~Xia, and X.~Mao, ``Cct5: A
  code-change-oriented pre-trained model,'' in {\em Proceedings of the 31st ACM
  Joint European Software Engineering Conference and Symposium on the
  Foundations of Software Engineering}, pp.~1509--1521, 2023.

\bibitem{wang2024divide}
S.~Wang, B.~Lin, L.~Chen, and X.~Mao, ``Divide-and-conquer: Automating code
  revisions via localization-and-revision,'' {\em ACM Transactions on Software
  Engineering and Methodology}, 2024.

\bibitem{Wei2023}
J.~Wei, X.~Wang, D.~Schuurmans, M.~Bosma, B.~Ichter, F.~Xia, E.~Chi, Q.~Le, and
  D.~Zhou, ``Chain-of-{Thought} {Prompting} {Elicits} {Reasoning} in {Large}
  {Language} {Models},'' Jan. 2023.
\newblock arXiv:2201.11903 [cs].

\bibitem{Min2022}
S.~Min, X.~Lyu, A.~Holtzman, M.~Artetxe, M.~Lewis, H.~Hajishirzi, and
  L.~Zettlemoyer, ``Rethinking the {Role} of {Demonstrations}: {What} {Makes}
  {In}-{Context} {Learning} {Work}?,'' Oct. 2022.
\newblock arXiv:2202.12837 [cs].

\bibitem{Nashid2023}
N.~Nashid, M.~Sintaha, and A.~Mesbah, ``Retrieval-{Based} {Prompt} {Selection}
  for {Code}-{Related} {Few}-{Shot} {Learning},'' in {\em 2023 {IEEE}/{ACM}
  45th {International} {Conference} on {Software} {Engineering} ({ICSE})},
  (Melbourne, Australia), pp.~2450--2462, IEEE, May 2023.

\bibitem{Liu2023}
C.~Liu, X.~Bao, H.~Zhang, N.~Zhang, H.~Hu, X.~Zhang, and M.~Yan, ``Improving
  {ChatGPT} {Prompt} for {Code} {Generation},'' May 2023.
\newblock arXiv:2305.08360 [cs].

\bibitem{White2023b}
J.~White, S.~Hays, Q.~Fu, J.~Spencer-Smith, and D.~C. Schmidt, ``{ChatGPT}
  {Prompt} {Patterns} for {Improving} {Code} {Quality}, {Refactoring},
  {Requirements} {Elicitation}, and {Software} {Design},'' Mar. 2023.
\newblock arXiv:2303.07839 [cs].

\end{thebibliography}

\end{document}